\documentclass[submission,copyright,creativecommons]{eptcs}
\providecommand{\event}{PROLE 2016}
\usepackage{paralist}
\usepackage{graphics}
\usepackage{graphicx}
\usepackage{hyperref}
\usepackage{subfigure}
\usepackage{url}
\usepackage{underscore}
\usepackage{amsthm}
\usepackage{color}

\newtheorem{example}{Example}[section]

\title{Comparing MapReduce and {\em Pipeline} Implementations for Counting Triangles}
\author{Edelmira Pasarella\thanks{This research is supported in part by funds from the Spanish Ministry for Economy and Competitiveness (MINECO) and the European Union (FEDER funds) under grant COMMAS (ref. TIN2013-46181-C2-1-R)} 
\institute{Computer Science Department\\ Universitat Polit\`ecnica de Catalunya\\ 
 Barcelona, Spain}
\email{edelmira@cs.upc.edu}
\and
Maria-Esther Vidal\thanks{This research is partially supported by the European Union Horizon 2020 programme for the project BigDataEurope (GA 644564).}
\institute{Fraunhofer IAIS\\ Bonn,  Germany }
\institute{Universidad Sim\'on Bol\'ivar\\Caracas, Venezuela}
\email{vidal@cs.uni-bonn.de}
\and
Cristina Zoltan
\institute{Computer Science Department\\ Universitat Polit\`ecnica de Catalunya\\ 
 Barcelona, Spain}
 \institute{Universidad Internacional de Ciencia y Tecnolog\'ia\\ Panam\'a }
\email{zoltan@cs.upc.edu}
}
\def\titlerunning{Comparing MapReduce and Pipeline Implementations for Counting Triangles}
\def\authorrunning{E. Pasarella, M.E. Vidal \& C. Zoltan}
\begin{document}
\maketitle

\begin{abstract}
A common method to define a parallel solution for a computational problem consists in finding a way to use the  {\it Divide \& Conquer} paradigm in order to have processors acting on its own data and  scheduled in a parallel fashion.  MapReduce is a programming model that follows this paradigm, and allows for the definition of efficient solutions by both {\it decomposing} a problem into steps on subsets of the input data and {\it combining} the results of each step to produce final results.  Albeit used for the implementation of a wide variety of computational problems, MapReduce performance can be negatively affected whenever the {\it replication factor} grows or the {\it  size of the input} is larger than the resources available at each processor. 
In this paper we show  an alternative approach to implement the {\it Divide \& Conquer} paradigm, named {\it dynamic pipeline}. The main features of {\it dynamic pipelines} are illustrated on a parallel implementation of the well-known problem of counting triangles in a graph. This problem is especially interesting either when the input graph does not fit in memory or is  dynamically generated.  To evaluate the properties of {\it pipeline}, a dynamic {\it pipeline} of processes and an {\it ad-hoc} version of  MapReduce are implemented in the language Go, exploiting its ability to deal with channels and spawned processes. An empirical evaluation is conducted on graphs of different topologies, sizes, and densities. Observed results suggest that {\it dynamic pipelines} allows for an efficient implementation of the problem of counting triangles in a graph, particularly, in dense and large graphs, drastically reducing the execution time with respect to the MapReduce implementation. 
\end{abstract}
\section{Introduction}

The  \textit{Divide \& Conquer} paradigm~\cite{Atallah:2010:ATC:1882757}  is an algorithm design schema that enables to solve large and complex  computational problems in three stages: 
\begin{inparaenum}[\itshape i\upshape)]
\item  {\it Divide}:  an instance of the problem  is partitioned into subproblems; 
\item {\it Conquer}:  the subproblems are solved independently;
\item  {\it Combine}: the solutions of the subproblems are combined to produce the final results.  
\end{inparaenum}
The  \textit{Divide \& Conquer} paradigm  is well-known for giving good complexity results.
MapReduce \cite{opac-b1134500}  is  an implementation schema/programming paradigm  of the \textit{Divide \& Conquer} paradigm,  extensively used in the implementation of complex problems.  The  {\it divide} stage  is done by establishing an {\it  equivalence  relation}  on the set of values which are the images of an input set transformed by a {\it mapping process},   such that  in the {\it Conquer}  stage {\it reducers}  can act on disjoint sets,  i.e., each {\it reducer} acts on a different equivalence class. Finally, other processes  collect  the {\it partial results} produced by the reducers to generate the solution in the {\it combine} stage. 

Frameworks that implement the MapReduce schema have had  great success and are mostly addressed to run on distributed architectures. 
Parallelism is a mean for speeding up solutions for computational programs with large amounts of data in memory and that have, in general, a regular behavior. 
The MapReduce scheme utilizes  the Valiant's Bulk Synchronous Parallel (BSP) model of computation~\cite{Valiant:1990:BMP:79173.79181}, and it is defined in terms of a pipe of three stages: \textit{Map}, \textit{Shuffle}, and \textit{Reduce}\footnote{Some implementations combine the shuffle step with the Map step.}. Map transforms a domain, where the {\it equivalence relation} can be established. Shuffle divides the collection into {\it sub-collections} where the reducers can act independently; 
the communication between processes in the pipe is done via distributed files which act as shared memory for  the processors. Some problems require the composition  of several  MapReduce processes. The number of composed processes is called the {\it number of passes} the solution requires.   MapReduce implementations require that users provide at least the code for the \textit{Map} and for \textit{Reduce} processes, as well as determine the {\it number of processors} assigned to the solution. Hadoop~\cite{White:2009:HDG:1717298} is a framework that provides a programing file system and operating system abstractions for distributing data and processing. It also enables the evaluation and testing of MapReduce implementations, and can recover itself from system failures. 
The success of the MapReduce schema for solving problems having massive input data has been extensively reported in the literature~\cite{DBLP:series/synthesis/2010Lin}, however, it is also known  the MapReduce approach is not suitable for solving problems that require the existence of a shared global state at execution time and solutions that require  several passes.  

In this work, we tackle limitations of the MapReduce programing schema, and present an alternative computing approach of the \textit{Divide \& Conquer} paradigm  for solving problems with massive input data. This implementation is based on  a {\it dynamic  pipeline} of processes via an asynchronous model of computation, synchronized by channels. 
 A \textit{dynamic pipeline} is  like an ordinary pipeline, but the number of stages is not fixed and are dynamically created  at runtime, i.e., \textit{dynamic pipeline} is able to adapt itself to the characteristics of a problem instance. 
  
  To be concrete,  we consider  the problem of triangle counting. This problem is relevant for a wide variety of problems in graph data analytics, query optimization, and graph partitioning, e.g., counting triangles represents a building block for computing the clustering coefficient of a graph. We present an implementation of counting triangles based on two rounds of the MapReduce schema presented by Suri and Vassilvitskii \cite{Suri:2011:CTC:1963405.1963491},  and a {\it dynamic pipeline} implementation following the approach proposed by Ar\'aoz and Zoltan~\cite{DBLP:journals/corr/AraozZ15}; features of the Go  programming language \cite{GO} are used to provide efficient implementations of these approaches. 
  
We empirically evaluate the performance of the {\it dynamic pipeline} and MapReduce based implementations on a large variety of graphs of different size and density. The observed results suggest that the {\it dynamic pipelining} implementation outperforms the MapReduce based  solution for dense graphs and with a large number of edges; savings in execution time can be of up to two orders of magnitude. 

In summary in this paper, we make the following contributions:
\begin{itemize}
    \item A comparison of the MapReduce and {\it dynamic pipeline} programing schemas in the resolution of the problem of counting triangles.
    \item Implementations  in the Go language of two algorithms that follow the MapReduce and {\it dynamic  pipeline} programing schemas to solve the  problem of counting triangles. These algorithms exploit the main properties of Go, i.e., channels and spawned processes, and correspond to implementations of the {\it dynamic  pipeline} and MapReduce programming schemas under the same conditions.
    \item An empirical evaluation  of the MapReduce and {\it dynamic pipeline} based algorithms to evaluate the performance of both algorithms in a variety of graphs of different density, topology, and size.  
\end{itemize}

This paper is a revised and extended version of a short paper presented at the Alberto Mendelzon Workshop on Foundations of Data Management (AMW2016) \cite{AMW2016}. The work is organized as follows: Section \ref{preliminaries} describes the problem of counting triangles, and the main features of the Go programing language, while Section \ref{solutions} presents the implementations of the problem of  triangle counting in both MapReduce and {\it dynamic pipeline}  using the Go language. In Section \ref{Experiments}, results of the experimental study are reported and discussed. Finally, we present the concluding remarks and future work in Section \ref{conclude}. 
\section{Preliminaries}
\label{preliminaries}
\subsection{The Problem of Counting Triangles}\label{the problem}

The problem of counting triangles in a graph has a simple formulation: Count the number of distinct set of  3 edges taken from a given graph, such that  $\{(a,b), (b,c), (c,a)\}$, i.e., the number of complete subgraphs of three nodes of the given graph~\cite{AlonYZ97}.
Counting triangles is a {\it building block} for determining the connectivity of a community around a node, representing a relevant problem in the context of network analysis.  Specifically, given the size of existing networks, efficiency needs to be ensured, and existing approaches exploit the benefits of parallel computation in  MapReduce~\cite{DBLP:journals/cse/Cohen09,DBLP:journals/ipl/PaghT12,Suri:2011:CTC:1963405.1963491}, while others implement approximate solutions to the problem~\cite{DBLP:conf/pods/BuriolFLMS06,DBLP:conf/swat/KutzkovP14a,DBLP:conf/bigdataconf/RahmanH13}. Parallel MapReduce approaches follow the MapReduce programming schema for efficiently counting triangles; however, intrinsic limitations of the MapReduce programming schema may prevent these approaches from scaling up to large and dense graphs.  On the other hand, approximation algorithms rely on estimators of the numbers of edges in a graph to approximate the number of triangles. Nevertheless, as shown by Bar-Yossef et al. \cite{Chu:2012:TLM:2382577.2382581}, theoretical bounds suggest that is impossible to precisely approximate these numbers in general graphs efficiently. Recently, Hu et. al~\cite{HuQT15a,HuTC14} propose efficient algorithms that rely on specific graph representations, e.g., adjacent lists. Albeit effective, these algorithms do not follow the MapReduce programming schema,  and they will require  a pre-processing phase to generate internal representations of a graph.

We tackle an {\it exact solution} to the problem, and present two algorithms that exploit the properties of MapReduce and pipeline in the Go programing language. To be concrete, we present Go implementations for two algorithms: a) the one  proposed by Suri and Vassilvitskii~\cite{Suri:2011:CTC:1963405.1963491}; b) the algorithm proposed by Ar\'aoz and Zoltan~\cite{DBLP:journals/corr/AraozZ15} where  a graph is represented as a sequence of unordered edges. The  algorithm by Ar\'aoz and Zoltan can be naturally extended to a triangle listing algorithm. As a precondition, the problem of counting triangles receives undirected simple graphs, i.e.,  no multiple edges are admitted; to ensure this requirement multiple edges are  filtered in a pre-processing stage. 

\subsection{Main features of the Go programming language:  Channels and Goroutines}
Go~\cite{GO} is a programming language that  facilitates efficient implementations of parallel programs, and naturally supports concurrency, as well as
processes for automatic memory management and garbage collection.
Additionally, Go makes available \textit{goroutines} which are lightweight threads managed by Go during runtime. Goroutines are needed not only for dynamically spawning processes, but for describing processes that resume their work (retaining all the values) when stop being blocked. 

Go  also provides a mechanism of channels to communicate concurrent goroutines, and pass values from senders to receivers.
Receivers always block until there is data to receive. If the channel is unbuffered, the sender blocks until the receiver has received the value. If a channel has a buffer, the sender blocks only until the value has been copied to the buffer; while a buffer is full, this means waiting until some receiver has retrieved a value.
The Go model of computation using channels, is a synchronous message passing. Task parallelism is obtained by the rule that all unblocked processes can run in parallel. 

These features make Go a suitable programing language for the problem of counting triangles, and enable efficient implementations of both \textit{dynamic pipeline} and MapReduce approaches.
Figures \ref{topologiaSolucion2} and \ref{topologiaSolucion1} illustrate the structure of the two-round MapReduce solution and the {\it dynamic pipeline} approach for the studied problem, respectively. Both implementations are explained in next section.

\section{Solutions to the Problem of Counting Triangles} 
\label{solutions}

Suri  and  Vassilvitskii~\cite{Suri:2011:CTC:1963405.1963491}  present a composition of  two  MapReduce algorithms for solving the triangle counting problem. In the first  MapReduce application,  the input is a set of edges of an undirected graph, and the output is a set of 2-length paths having a given responsible node. Each 2-length path with its responsible node is represented by a triple (path-triple). The second application of MapReduce  receives the path-triples generated by the previous application of MapReduce and the edge-triples, i.e., the edges present in the original input graph with an empty middle element. For each triple, the pair of its end nodes is used as its key. The reducer task identifies if  a path-triple and an edge-triple are in the same cluster. If so, the number of triangles is equal to the cluster size minus one; otherwise, the number of triangles in the cluster is zero. Adding the number of triangles in each cluster gives the total number of triangles in the graph.  It is common that the number of reducers coincides with the number of available processors. So the behavior is not smooth
in the number of processors.

\begin{figure}[h!]
\begin{center}
		\includegraphics[width=0.7\textwidth]{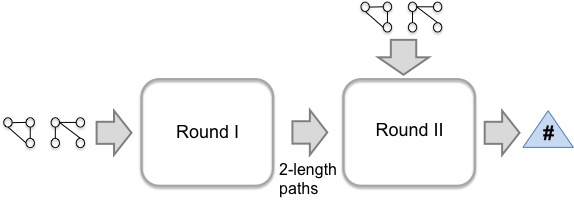}
\end{center}
\caption{{\bf MapReduce algorithm proposed in \cite{Suri:2011:CTC:1963405.1963491}.} A two-round MapReduce algorithm for counting triangles. Round I generates paths of length 2, while during Round II,  edges of the input graph and paths of length 2 are used to count the number of triangles in the input graph  }
\label{fig:2rounds}
\end{figure}

\subsection{A MapReduce solution}
\begin{figure}[h!]
\begin{center}
		\includegraphics[width=0.8\textwidth]{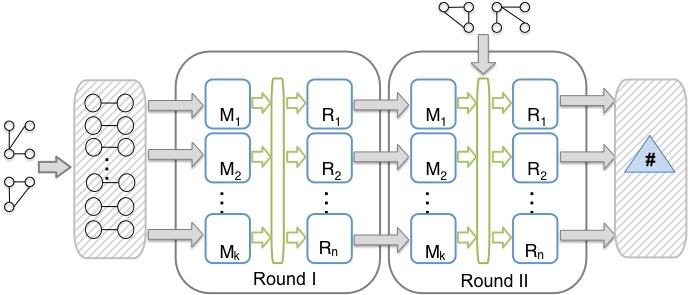}
\end{center}
\caption{{\bf  MapReduce for Triangle Counting}. Go implementation of the two-round of MapReduce  algorithm in Figure \ref{fig:2rounds}.  Small rounded boxes represent mappers and reducers, while large rounded boxes correspond to rounds. Grey and clear arrows represent I/O operations and channel communications, respectively. Mappers and reducers communicate via a channel array. During Round I, edges in the input graph are partitioned to feed the mappers, and reducers generate 2-length paths. In Round II, 2-length paths and edges feed the reducers to count the number of triangles}
\label{topologiaSolucion2}
\end{figure}
\begin{figure}[h!]
\begin{center}
		\includegraphics[width=0.33\textwidth]{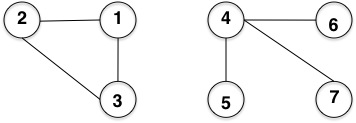}
\end{center}
\caption{{\bf Running Example.} A graph including only one triangle is used as running example }
\label{fig:inputgraph}
\end{figure}
\paragraph{MapReduce Implementation:} Figure \ref{topologiaSolucion2} shows the phases of our implementation of  Suri  and  Vassilvitskii's  algorithm \cite{Suri:2011:CTC:1963405.1963491}. The program receives as an input a file which is partitioned into as many files as 
the number of mappers, e.g., the number of available cores. In order to reduce the execution time in the MapReduce implementation, hashing is applied during the Map stage and the mappers communicate via buffered channels with the reducers. The output of the round I is the set of 2-length paths which are sent to files. In round II, these paths are merged with the input graph edges, and distributed to the reducers. 

The output of each reducer is the number of triangles found in its input, i.e., triangles formed by 2-length paths having the same end points and connected by an edge. A process collects the outputs from the reducers to give the final result. 
Our implementation follows the \textit{MapReduce Online} approach proposed by Condie et. al. \cite{Condie:2010:MO:1855711.1855732}, and avoids blocking communication between stages.  
\begin{figure}[h!]
\begin{center}
\subfigure[Shuffle Phase: adjacent lists are enumerated]{
\includegraphics[width=0.43\textwidth]{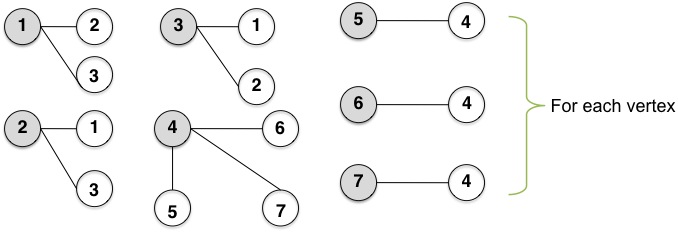}
\label{fig:r1-shuffle}
}  \quad
\subfigure[Reduce Phase: 2-length paths are enumerated] {
\includegraphics[width=0.5\textwidth]{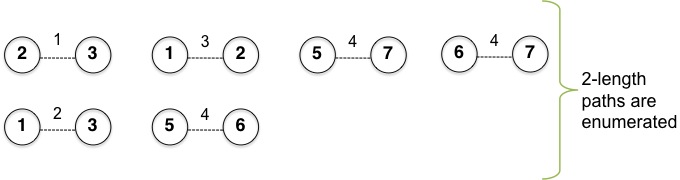}
\label{fig:r1:reduce}
}
\end{center}
\caption{{\bf Example of Round I.} Execution of the MapReduce algorithm on the graph of Figure \ref{fig:inputgraph}. a) Adjacent lists are enumerated during the the Shuffle phase; b) Reducers enumerate paths of length 2}
\label{fig:MR-round1}
\end{figure}
\begin{figure}[h!]
\begin{center}
\subfigure[Shuffle Phase: Triples are grouped]{
\includegraphics[width=0.30\textwidth]{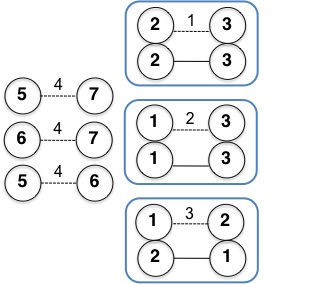}
\label{fig:r2-shuffle}
}  \quad
\subfigure[Reduce Phase: triangles are counted] {
\includegraphics[width=0.42\textwidth]{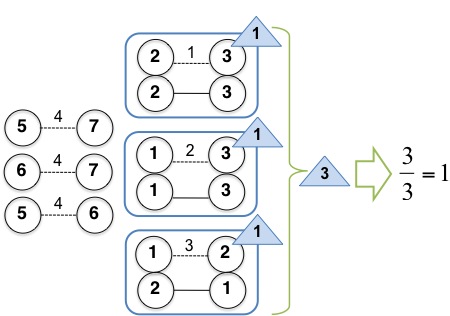}
\label{fig:r2:reduce}
}
\end{center}
\caption{{\bf Example of Round II.} Execution of the MapReduce algorithm on  graph in Figure \ref{fig:inputgraph}. a) Groups of triples are produced during the Shuffle phase; and b) Reducers count triangles in groups of triples}
\label{fig:MR-round2}
\end{figure}

\begin{example}
Consider the graph shown in Figure \ref{fig:inputgraph} where the input to the algorithm is given by the sequence of edges $(2,1), (1,3),(4,5), (2,3), (4,7), (4,6)$. Figure \ref{fig:MR-round1} shows the result at the end of round I, where mappers keep the edges without change, while for each node, the shuffle process produces a set of edges incident to each node i.e., (1,[2,3]), (2,[1,3]), (3,[1,2]), (4,[5,6,7]), (5,[4]), (6,[4]) and (7,[4]) as seen in Figure \ref{fig:r1-shuffle}. Each reducer produces paths of length 2; middle nodes of these paths are presented as labels in the edges as shown in Figure \ref{fig:r1:reduce}. In the running example, 6 paths of length 2 are produced. 

During round II, the mappers transform the edges into triples with an empty middle node. Afterwards, the shuffle groups triples (with a center node or without it) having the same end points as seen in Figure \ref{fig:r2-shuffle}.  A reducer takes the set of values and if  the set includes a path of length 2 and one edge, it will  give as a result the size of the set minus one; otherwise, the result is 0. We use the notation (a,x,b) to describe a path of length 2 from a to b or an edges from a to b, then  Figure \ref{fig:r2:reduce} shows that reducers with input sets that have elements of the form (2,x,3), (1,x,2), and (1,x,3) will output 1 triangle, while the ones having as input (5,x,7), (6,x,7), or (5,x,6) will report 0 triangles. Each triangle is reported 3 times as we can see in Figure \ref{fig:r2:reduce}; therefore, the total sum provided by the reducers needs to be divided by 3.
\color{blue}
\end{example}

\subsection{A Pipeline solution}
\begin{figure}[tb]
\centering
\includegraphics[width=0.7\textwidth]{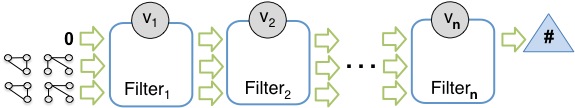}
\caption{{\bf Pipeline Topology for Triangle Counting}. Dynamic composition of filters that work on set of values not consumed by a previous filter. Rounded boxes represent filters, and grey circles and clear arrows correspond to responsible nodes and data flow, respectively. Filters have three input and three output channels represented by unfilled arrows. The first filter receives the input graph in the second and third channel. The first channel carries the number of triangles; initially, this value is zero. Filters are dynamically created at runtime and executions are adapted to the characteristics of the input}
\label{topologiaSolucion1}
\end{figure}
The pipeline solution is a dynamic composition of filters specialized to nodes of the input graph, and each one works on a set of values not consumed by a previous filter. Each filter has three inputs and three outputs for receiving/sending  messages from/to their neighbors.  The program structure is a pipe of processes. Initially the pipe is empty, and during execution, the pipe grows/shrinks based on data flowing in the pipe. The first filter is created using the first incoming edge. The first filter will receive the complete set of edges, using the third input.  When created, each filter specializes itself with the first incoming  edge, using the first node of the edge as \textit{responsible node} and adding the other to an adjacent list.  Afterwards, each filter treats the  incoming edges, keeping those incident to its  responsible node  and sending the others  to its neighbor. 
Filters are created dynamically, as new values  not consumed  by already created filters arrive. As each filter consumes at least one value, no more than  $| V|-1 $ filters can be created. It is an upper bound because the graph does not have isolated nodes. The number of filters is equal to the number of classes generated by the relations on  the original set. The partition relation on edges  is created during the execution using responsible nodes as representatives of the set of edges adjacent with the responsible node. The set of responsible nodes is a \textit{dominator set}. Given a graph $G=(V,E)$, a \textit{dominator set} $S$ of $G$ is a subset such that every node 
$n \in V-S$ is an adjacent node of a node $k \in S$.  
Whenever there are no more edges in the third input, the filter enters into a second phase. It counts the number of edges flowing in the second input having both endpoints in the adjacency list of  the filter responsible node. If there are no more edges flowing in the second input, each filter has the number of triangles, having the responsible node as one of its nodes. Therefore, each triangle is counted only once. Then, the filter enters in the third phase where it outputs the number of triangles already counted and dies; at this point, the pipeline shrinks. The first channel of each filter is used to collect the total number of triangles in the graph.
A proof of correctness of this \textit{pipeline} algorithm can be found in \cite{DBLP:journals/corr/AraozZ15}.

\begin{figure}[h!]
\begin{center}
	\subfigure[Two filters are created]{
		\includegraphics[width=0.24\textwidth]{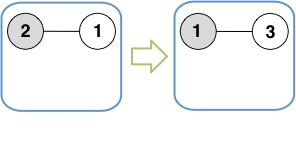}
		\label{fig:ex-pl1}
	}  \quad
	\subfigure[Partition of the edges is generated] {
		\includegraphics[width=0.4\textwidth]{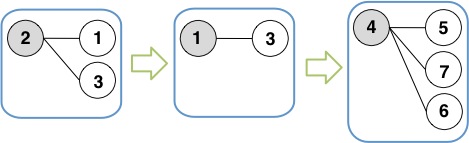}
		\label{fig:ex-pl2}
	}
\end{center}
\caption{{\bf  Dynamic Pipeline Partition phase}. Execution of the partitioning phase of the pipeline algorithm. Filters are created according to the input edges.  Responsible nodes are represented using grey circles, while filters are modeled using rounded boxes. Arrows represent data flow through the pipeline}
\label{fig:pipeline-ex-phase1}
\end{figure}

\begin{figure}[h!]
\begin{center}
	\subfigure[Edges (2,1) and (1,3) have both been processed by the left-most filter, while the middle filter has processed the edge (2,1)]{
		\includegraphics[width=0.33\textwidth]{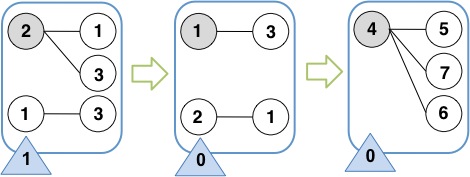}
		\label{fig:ex-pl3}
	}  \quad
	\subfigure[Edge (4,5), (1,3), and (2,1) have all been processed by the left-most filter; (1,3) and (2,1) have both been processed by the middle filter; while the right-most filter has processed the edge (2,1)] {
		\includegraphics[width=0.33\textwidth]{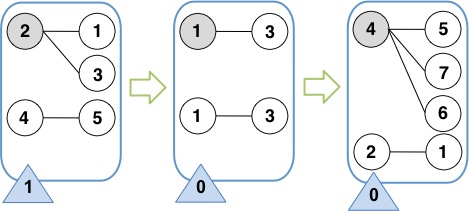}
		\label{fig:ex-pl4}
	}
\end{center}
\caption{{\bf Pipeline counting triangle execution-Counting phase}. All triangles adjacent to their corresponding responsible node are counted. Responsible nodes are represented using grey circles, while created filters are modeled using rounded boxes. Arrows represent data flow through the dynamic pipeline}
\label{fig:pipeline-ex-phase2}
\end{figure}

\begin{figure}[h!]
\begin{center}
	\subfigure[Each filter receives the number in triangles in previous filters, adds it to its triangles' account and deactivates itself]{
		\includegraphics[width=0.37\textwidth]{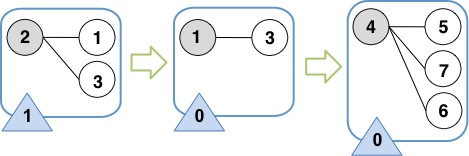}
		\label{fig:ex-pl5}
	}  \quad
	\subfigure[The first filter is no longer active and the second one holds the partial count] {
		\includegraphics[width=0.25\textwidth]{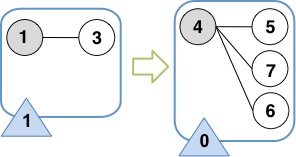}
		\label{fig:ex-pl6}
	}
\end{center}
\caption{{\bf Pipeline counting triangle execution-Aggregation phase}. Elasticity of the dynamic pipeline is illustrated. Responsible nodes are represented using grey circles, while created filters are modeled using rounded boxes. Arrows represent data flow through the dynamic pipeline}
\label{fig:pipeline-ex-phase3}
\end{figure}
The following example illustrates how the algorithm proceeds.

\begin{example}
Let us consider again the graph shown in Figure \ref{fig:inputgraph}, where the input to the algorithm is given by the sequence of edges $(2,1), (1,3),(4,5), (2,3), (4,7), (4,6)$. Figures \ref{fig:ex-pl1} and  \ref{fig:ex-pl2} show 
the state of the algorithm  in the Partition phase. Figure \ref{fig:ex-pl1} presents the state when the input is partially consumed, while Figure \ref{fig:ex-pl2} presents the state after reading and processing all the edges. In Figure \ref{fig:ex-pl1}, edges (2,1) and (1,3) are processed and only two filters are created with responsible nodes 2 and 1. At the end of this phase, as shown in Figure \ref{fig:ex-pl2} there are three filters with corresponding responsible nodes: 2, 1, and 4. Further, each filter keeps all the adjacent nodes to the corresponding responsible node. Figure \ref{fig:pipeline-ex-phase2} gives snapshots of the state of the algorithm in the Counting phase. At the beginning of this phase, each filter keeps the nodes adjacent to the corresponding responsible node, not consumed by previous fiters. The edges flow in the pipe, and in Figure \ref{fig:ex-pl3}, we see that $(2,1), (1,3)$ are being processed, and as a result the filter with responsible node 2 is able to count a triangle (\includegraphics[width=0.032\textwidth]{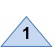}), while the second filter does not count any triangle. In Figure \ref{fig:ex-pl4} the edges $(2,1), (1,3),(4,5)$ are processed by the filters, none of them changing the number of triangles having their responsible node as one of its nodes. Figure  \ref{fig:pipeline-ex-phase3} shows states of phase 3, where the partial count on each filter is transmitted to its neighbor  in order to collect the total sum. In Figure \ref{fig:ex-pl4} we can see that after transmitting its triangle count to its neighbour, the first filter dies and the one with responsible node 1, holds the partial triangle count. 
\end{example}

\paragraph{Pipeline Implementation:}\label{pipe implementation} Figure \ref{topologiaSolucion1} shows the topology of our implementation of the \textit{pipeline} of the algorithm proposed by Ar\'aoz and Zoltan~\cite{DBLP:journals/corr/AraozZ15}. In particular, this topology is the composition of filters specialized to  nodes of the input graph, and each one works on a set of values that are not consumed by the previous filter. The first filter receives the complete set of edges on the third input. Each new filter specializes itself with the first incoming edge, using the first node of the edge as responsible node and adds the other one to an adjacent list. Afterwards, each filter treats the incoming edges, keeping those nodes belonging to edges incident to its responsible node and sending the others to their neighbor. 

Whenever all the edges are being processed, each filter has the nodes of the received edges incident to its responsible node.
The number of filters is equal to the number of classes generated by the relation on the original set.
Filters are processes/\textit{goroutines} that communicate via unbuffered channels and each process is specialized by a responsible node. \textit{Goroutines} have three input channels and three output channels. Processes use lists to keep nodes adjacent to the responsible one. 
In each filter, every incoming edge in the second input is checked if it is incident to two nodes adjacent to the corresponding responsible one. If so, the number of triangles found is increased by one. When there are no more edges, each filter has the total number of triangles in the graph that includes its responsible node. The first channel will carry the number of triangles found by each \textit{goroutines}. A final process adds up the partial results.
\section{Experiments}\label{Experiments}
In this section, we present the experimental results of MapReduce and Pipeline implementations for the problem of  counting triangles. The goal of the experiment is to analyze the impact of graph properties on time and space complexity of both implementations. We study the following research questions: 
\begin{inparaenum}[\bf {\bf RQ}1\upshape)]
\item Is the Pipeline based implementation able to overcome the {\it performance} of MapReduce implementation independently of the input graph characteristics?;
\item Are {\it density}, {\it topology}, and {\it size} of the input graph equally affecting Pipeline and  MapReduce implementations?;
\item Is the {\it number of cores} equally affecting Pipeline and  MapReduce implementations?.
\end{inparaenum}
The experimental configuration to evaluate these research questions is as follows:
~\\
\noindent

\begin{table}[htbp]
\centering
\begin{tabular}{|l|r|r|r| r|}
\hline
\multicolumn{1}{|c|}{\textbf{Graph}} & \multicolumn{1}{c|}{\textbf{\# Vertices}} & \multicolumn{1}{c|}{\textbf{\# Arcs}} & \multicolumn{1}{c|}{\textbf{Density}}  & \multicolumn{1}{c|}{\textbf{File size}} \\ \hline
\textsf{DSJC.1} & 1,000 & 99,258 & {\bf 0.10} & 1.1MB\\ \hline
\textsf{DSJC.5} & 1,000 & 499,652 &  {\bf 0.50} &5.2MB\\ \hline
\textsf{DSJC.9} & 1,000 & 898,898 &  {\bf 0.90} &9.3MB\\ \hline
\textsf{Fixed-number-arcs-0.1(FNA.1)} & 10,000 & 10,000,000 &  {\bf 0.10} & 140MB\\ \hline
\textsf{Fixed-number-arcs-0.5 (FNA.5)} & 4,472 & 10,000,000 &  {\bf 0.50} & 138MB\\ \hline
\textsf{Fixed-number-arcs-0.9 (FNA.9)} & 3,333 & 10,000,000 &  {\bf 0.90} & 136MB\\ \hline
\textsf{USA-road-d.NY (NY)} & 264,346 & 733,846 &  {\bf 1.04E-5} & 13MB \\ \hline
\textsf{Facebook-SNAP(107)} & 1,911 & 53,498 & {\bf 1.47E-2} & 0.524MB \\ \hline
\end{tabular}
\hspace{1.5in}\parbox{0.8 \textwidth}{
\caption{{\bf Benchmark of Graphs} Graphs of different sizes and densities.  Density is defined as $\frac{\# Arcs}{\# Vertices*(\# Vertices-1)}$}}
\label{tab:graph}
\end{table}
~\\ 
{\bf Datasets:}  We compare these two implementations using graphs of different topologies, densities, and sizes.These graphs are part of the  9th DIMACS Implementation Challenge - Shortest Paths\cite{dimacs}; \textsf{\small DSJC.1}, \textsf{\small DSJC.5}, and \textsf{\small DSJC.9} are graphs with the same number of nodes and different densities, while in \textsf{\small Fixed-number-arcs-0.1(FNA.1)}, \textsf{\small Fixed-number-arcs-0.5(FNA.5)}, and \textsf{\small Fixed-number-arcs-0.9(FNA.9)}, the number of nodes is changed to affect the graph density.  \textsf{\small USA-road-d.NY} and \textsf{Facebook-SNAP(107)}\cite{Facebook} are real-world graphs that correspond to the New York City  road network and a Facebook subgraph, respectively.  Table 1 describes these graphs in terms of number of vertices, arcs,  graph density, and file size. 
~\\
\noindent
{\bf Metrics:} As evaluation metrics, we consider the execution time (ET) and Virtual-memory  (VM). ET  represents the elapsed time (in seconds) between the submission  of a job and completion of the job including the generation of the final results. VM represents the virtual memory consumed by the batch job measured in GB. Both ET and VM are reported by  the {\it qsub} command when a  batch job is submitted  to the machine~\cite{RdLab}. 
~\\
\noindent
{\bf Implementation:}
Programs are run on a node of the cluster of the RDLab-UPC\footnote{\url{https://rdlab.cs.upc.edu/}} having two processors Intel(R) Xeon(R) CPU X5675 of 3066 MHz with six cores each one. The configuration used in the experiments for submitting jobs is up to 12 cores and 40GB of RAM. Programs are implemented in Go 1.6~\cite{Go1}. The same job is executed 10 times and the average in reported, given enough shared memory and a timeout of five hours.

\begin{figure}[tb]
\centering
\includegraphics[height=5cm]{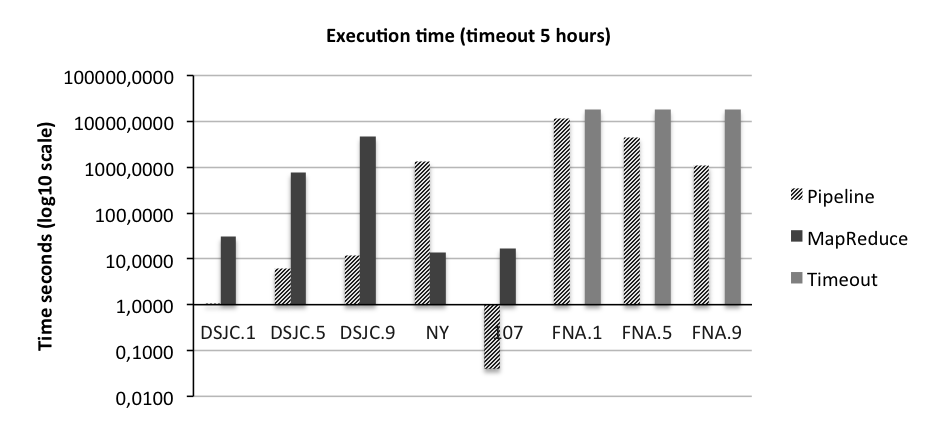}
\caption{{\bf Execution Time}. For graphs of different size and density,  performance of the Pipeline and MapReduce implementations is reported in terms of execution time (ET in  log10-scale secs). Jobs that time out at five hours are reported using light grey bars.}
\label{exp1}
\end{figure}

\begin{figure}[tb]
\centering
\includegraphics[height=5cm]{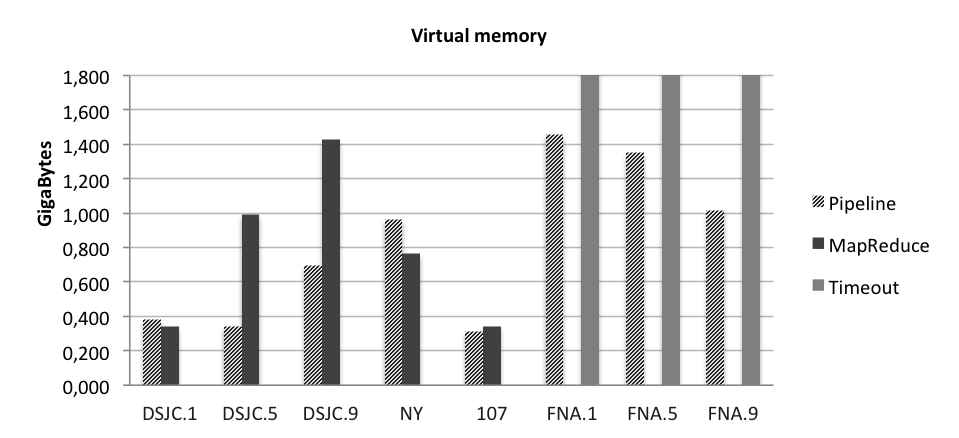}
\caption{{\bf Virtual Memory}. For graphs of different size and density,  performance of the Pipeline and MapReduce implementations is reported in terms of virtual memory (VM in GB). Jobs that time out at five hours are reported using light grey bars.}
\label{exp2}
\end{figure}

\noindent
{\bf Discussion}
Graphs with different  sizes and densities (0.10, 0.50, and 0.90) are evaluated to study our research questions {\bf RQ1} and {\bf RQ2}. Graphs with high density can be considered as the worst case for both program schemes. Plots in Figures~\ref{exp1} and \ref{exp2} report on execution time  (ET in log10-scale secs) and virtual memory (VM in GB) for each of the schemas. Jobs that time out at five hours are reported using grey bars. Jobs for the {\it pipeline} program in the different graphs are finished in less than 3 hours, while three jobs of the MapReduce implementations do not produce any response in five hours, i.e., these three jobs time out and are reported in light grey bars in Figures~\ref{exp1} and \ref{exp2}. 
The results suggest that the {\it pipeline} implementation exhibits the best results in response time and virtual memory consumption for graphs as the ones in \textsf{\small DSJC.1}, \textsf{\small DSJC.5}, \textsf{\small DSJC.9}, \textsf{\small FNA.1}, \textsf{\small FNA.5}, and \textsf{\small FNA.9}. Particularly, in the highly dense graphs, i.e., \textsf{\small DSJC.9} and  \textsf{\small FNA.9}, {\it pipeline} {\it drastically reduces} execution time with respect to  MapReduce. Similar performance is observed in the {\it real-world subgraph} of Facebook (Facebook-SNAP(107)), where {\it pipeline} execution time overcomes MapReduce by three orders of magnitude. Finally, the graph NY that represents the road network of NY city, is {\it highly sparse}  and the {\it pipeline} implementation generates a large number of processes that the Go scheduler is not able to deal with.

\begin{figure}[tb]
\centering
\includegraphics[height=5cm]{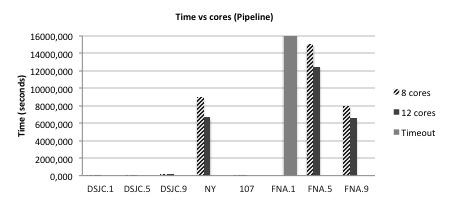}
\caption{{\bf Impact of the Number of Cores in Pipeline}.  For graphs of different size and density, the impact of the number of cores in  execution time (ET secs) of the pipeline implementation is reported. For graphs \textsf{\small DSJC.1}, \textsf{\small DSJC.5},  \textsf{\small DSJC.9}, and 107  execution time is less than 200 secs.}
\label{exp3}
\end{figure}

\begin{figure}[tb]
\centering
\includegraphics[height=5cm]{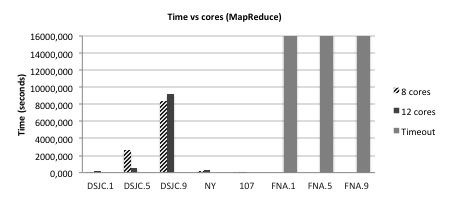}
 \caption{{\bf Impact of Number of Cores in MapReduce}.  For graphs of different size and density,  impact of the number of cores in execution time (ET secs) of the MapReduce  implementation is reported. Jobs that time out at five hours are reported using grey bars. Graphs \textsf{\small DSJC.1}, (Facebook-SNAP(107)) and NY, jobs of the MapReduce implementations produce the responses in less than 300 secs.}
\label{exp4}
\end{figure}

Our benchmark of graphs is also used to evaluate our research question {\bf RQ3}. Plots in Figures~\ref{exp3} and \ref{exp4} report on execution time (ET secs) for each of the schemas when the number of cores is eight or twelve. Jobs that time out at five hours are reported using light grey bars. For the graphs \textsf{\small DSJC.1}, \textsf{\small DSJC.5},  \textsf{\small DSJC.9}, and 107, jobs of the pipeline implementation requires less than 200 secs. to be completed and produce the response. Similarly, in graphs \textsf{\small DSJC.1}, (Facebook-SNAP(107)) and NY, jobs of the MapReduce implementations produce the responses in less than 300 secs. As the results reported in Figures~\ref{exp1} and \ref{exp2}, jobs for the MapReduce implementation time out at five hours for large graphs: \textsf{\small FNA.1}, \textsf{\small FNA.5}, and \textsf{\small FNA.9}. This negative performance of MapReduce is caused by the {\it replication factor} of the problem of counting triangles, i.e., the size of the set of 2-length paths (output in the first phase of MapReduce) is extremely large, O($n^2$) where $n$ is the number of graph vertices and these graphs have up to 10,000 vertices. 
These results corroborate our statement that the {\it pipeline} programming schema is a {\it promising} model for implementing complex problem and provides an adaptive solution to the characteristics of the input dataset. Furthermore,  {\it pipeline} is competitive with MapReduce and does not require any previous knowledge of the input dataset. 

\section{Conclusions and Future Work}\label{conclude}
We presented   an alternative approach, named {\it dynamic pipeline}, that follows the  \textit{Divide \& Conquer} paradigm, and relies on  a {\it dynamic  pipeline} of processes via an asynchronous model of computation for process communication. 
Users  of the {\it pipeline} approach need to provide a sequential code or {\it filters}, and require no understanding of standard concurrency mechanisms, e.g.,  threads and fine-grain concurrency control, which are aspects known to be difficult to deal with in order to obtain race condition free code in a parallel solution. Contrary to MapReduce, where implementations differ depending on the architectures, implementations based on the {\it pipeline}  approach can make transparent to users the implementation of channels. The  channel abstraction could have several concrete implementations 
that use shared memory, TCP pipes, or files temporarily persisted in a file system, e.g.,  as the ones provided by the Dryad distributed technologies \cite{Dryad}. This abstraction can allow for the deployment of the same program in a single machine with several cores, or a net of computing units. 

The well-known problem of  triangle counting  is utilized to illustrate the features of the {\it pipeline} approach as well as the differences with the MapReduce programming schema. Both  programs were implemented in multi-processor nodes. The proposed implementations provide exact solutions for counting triangles by exploiting the main characteristics of the Go programming language, i.e., the evaluation model, 
 a scheduler able to cope with dynamic scheduling, and  the notion of channels to enable the communication between processes. 
The performance of both implementations was empirically evaluated in artificial and real graphs with different sizes and densities. The observed results show a superiority in execution time for the pipeline schema even in dense graphs. The only case where  MapReduce exhibits a better performance corresponds to a graph where a large number of nodes have 
an approximate degree of 2, and this particular configuration results in a program that  negatively affects the Go scheduler. The results also suggest that the number of processors  has  a greater  positive impact on the pipeline schema than in  MapReduce. Based on these results, we can conclude that the {\it pipeline} approach is highly scalable, and is able to exhibit  performance gains on large problem instances with thousands of tasks, seeming to be  most promising  when a large number of processors  work on shared memory, e.g.,  in architectures as the one implemented in  \textit{The Machine}  from Hewlett Packard Labs\footnote{\url{http://www.labs.hpe.com/research/themachine/}}. 
In the future, we plan to continue the evaluation of the behavior of the {\it pipeline} approach in other complex computational problems, and create a programming framework.  Further, other algorithms for counting triangles in graph will be implemented and included in our evaluation study, e.g., algorithms by Hu et. al~\cite{HuQT15a,HuTC14}. However,  it is important to highlight that because these algorithms require different representations of a graph, e.g., adjacent lists, and are not implemented as MapReduce,  they will require  a pre-processing phase and will not be able to be used in graphs dynamically generated. In consequence, the experimental evaluation will have to be redefined in order to conduct a fair comparison of the studied approaches. 
\newline
\textbf{Acknowledgements.}  We would like to thank the reviewers for their insightful comments on the paper; all their comments led us to considerably improve our  work. Also, we thank  the staff of the \textit{Laboratori de Recerca i Desenvolupament} (RDlab) of the Computer Science Department of the UPC for their support  during execution of the experimental evaluation. 
\nocite{*}
\bibliographystyle{eptcs}
\bibliography{main.bbl}

 \end{document}